\documentclass[sigconf,balance=false,authorversion=true,nonacm=true]{acmart}
\usepackage{graphicx}
\usepackage{subcaption}

\AtBeginDocument{%
  }

\setcopyright{acmcopyright}
\copyrightyear{2023}
\acmYear{2023}
\acmDOI{XXXXXXX.XXXXXXX}

\acmConference[]{}{}
\begin{document}

%%
%% The "title" command has an optional parameter,
%% allowing the author to define a "short title" to be used in page headers.
\title{Discovering Image Usage Online: \\
A Case Study With ``Flatten the Curve''}

%%
%% The "author" command and its associated commands are used to define
%% the authors and their affiliations.
%% Of note is the shared affiliation of the first two authors, and the
%% "authornote" and "authornotemark" commands
%% used to denote shared contribution to the research.

\author{Shawn M. Jones}
% \authornote{Both authors contributed equally to this research.}
\email{smjones@lanl.gov}
\orcid{0000-0002-4372-870X}
\author{Diane Oyen}
\orcid{0000-0002-1353-3688}
% \authornotemark[1]
\email{doyen@lanl.gov}
\affiliation{%
  \institution{Los Alamos National Laboratory}
  \streetaddress{P.O. Box 1663}
  \city{Los Alamos}
  \state{New Mexico}
  \country{USA}
  \postcode{87545}
}

% \author{HIDDEN FOR REVIEW}
% \affiliation{%
%   \institution{HIDDEN FOR REVIEW}
%   \streetaddress{HIDDEN FOR REVIEW}
%   \city{HIDDEN FOR REVIEW}
%   \state{HIDDEN FOR REVIEW}
%   \country{HIDDEN FOR REVIEW}
%   \postcode{HIDDEN FOR REVIEW}
% }

\renewcommand{\shortauthors}{Jones \& Oyen}

\begin{abstract}
Understanding the spread of images across the web helps us understand the reuse of scientific visualizations and their relationship with the public. The ``Flatten the Curve'' graphic was heavily used during the COVID-19 pandemic to convey a complex concept in a simple form. It displays two curves comparing the impact on case loads for medical facilities if the populace either adopts or fails to adopt protective measures during a pandemic. We use five variants of the ``Flatten the Curve'' image as a case study for viewing the spread of an image online. To evaluate its spread, we leverage three information channels: reverse image search engines, social media, and web archives.  Reverse image searches give us a current view into image reuse. Social media helps us understand a variant's popularity over time. Web archives help us see when it was preserved, highlighting a view of popularity for future researchers. Our case study leverages document URLs can be used as a proxy for images when studying the spread of images online.
\end{abstract}

%%
%% The code below is generated by the tool at http://dl.acm.org/ccs.cfm.
%% Please copy and paste the code instead of the example below.
%%


%%
%% Keywords. The author(s) should pick words that accurately describe
%% the work being presented. Separate the keywords with commas.

%% A "teaser" image appears between the author and affiliation
%% information and the body of the document, and typically spans the
%% page.
% \begin{teaserfigure}
%   \includegraphics[width=\textwidth]{sampleteaser}
%   \caption{Seattle Mariners at Spring Training, 2010.}
%   \Description{Enjoying the baseball game from the third-base
%   seats. Ichiro Suzuki preparing to bat.}
%   \label{fig:teaser}
% \end{teaserfigure}

% \received{20 February 2007}
% \received[revised]{12 March 2009}
% \received[accepted]{5 June 2009}

%%
%% This command processes the author and affiliation and title
%% information and builds the first part of the formatted document.
\maketitle

\begin{figure}[ht]

    \begin{subfigure}[b]{0.2\textwidth}
        \centering
        \includegraphics[width=\textwidth]{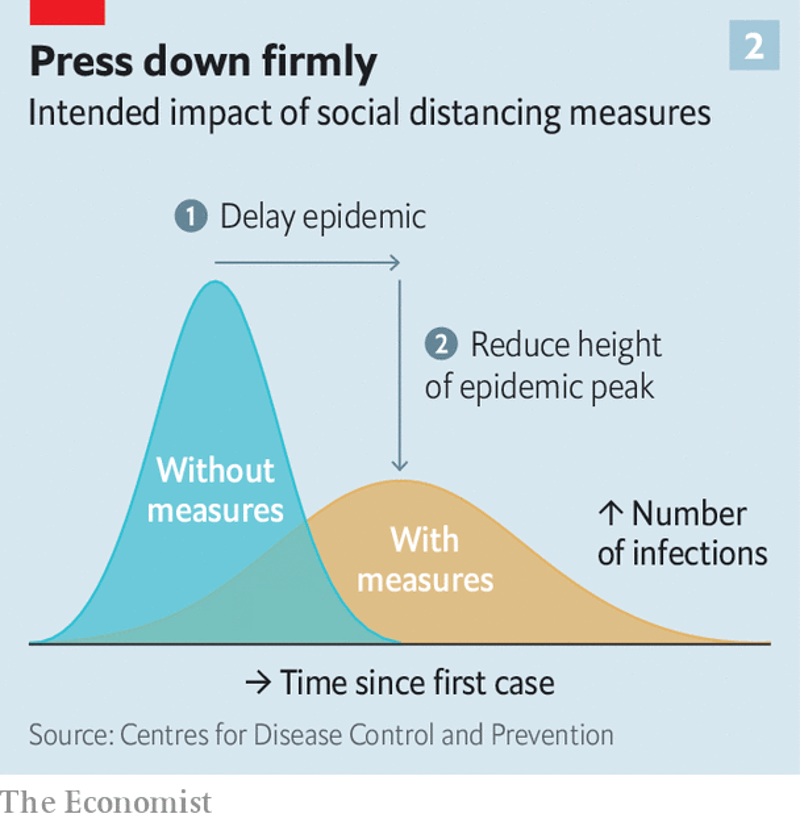}
        \caption{\emph{The Economist}}
        \label{fig:economist_ftc}
    \end{subfigure}%
    \begin{subfigure}[b]{0.2\textwidth}
        \centering
        \includegraphics[width=\textwidth]{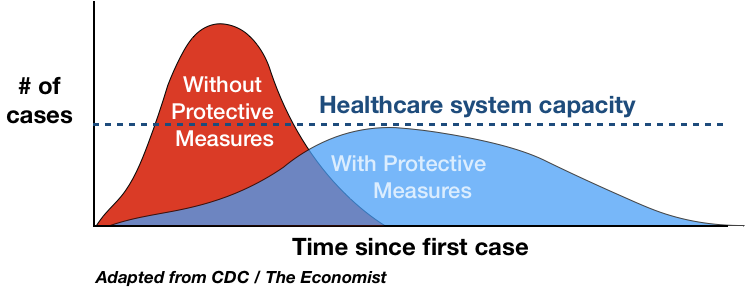}
        \caption{Drew Harris}
        \label{fig:harris_ftc}
    \end{subfigure}
    
    \begin{subfigure}[b]{0.2\textwidth}
        \centering
        \includegraphics[width=\textwidth]{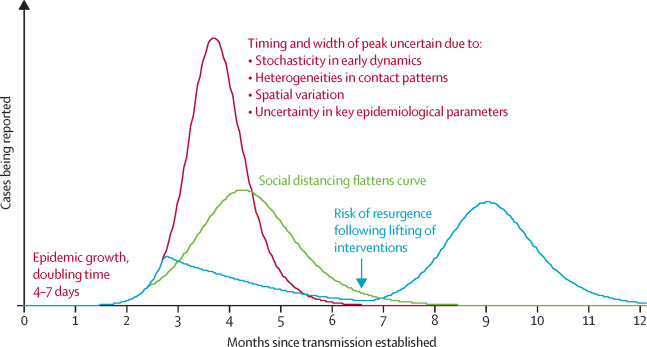}
        \caption{ \emph{The Lancet} \cite{lancet_2020} }
        \label{fig:lancet_ftc}
    \end{subfigure}%
    \begin{subfigure}[b]{0.2\textwidth}
        \centering
        \includegraphics[width=\textwidth]{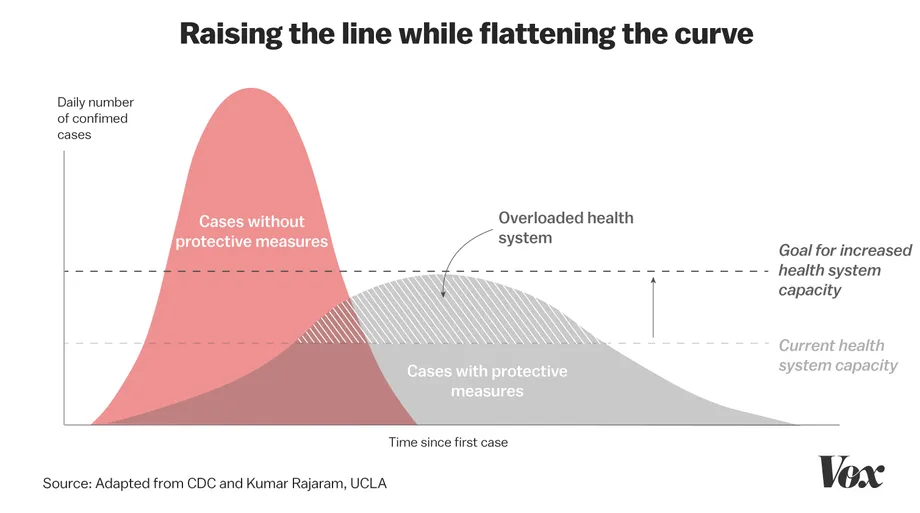}
        \caption{\emph{Vox} \cite{vox_2020}}
        \label{fig:vox_ftc}
    \end{subfigure}
    
    \begin{subfigure}[b]{0.2\textwidth}
        \centering
        \includegraphics[width=\textwidth]{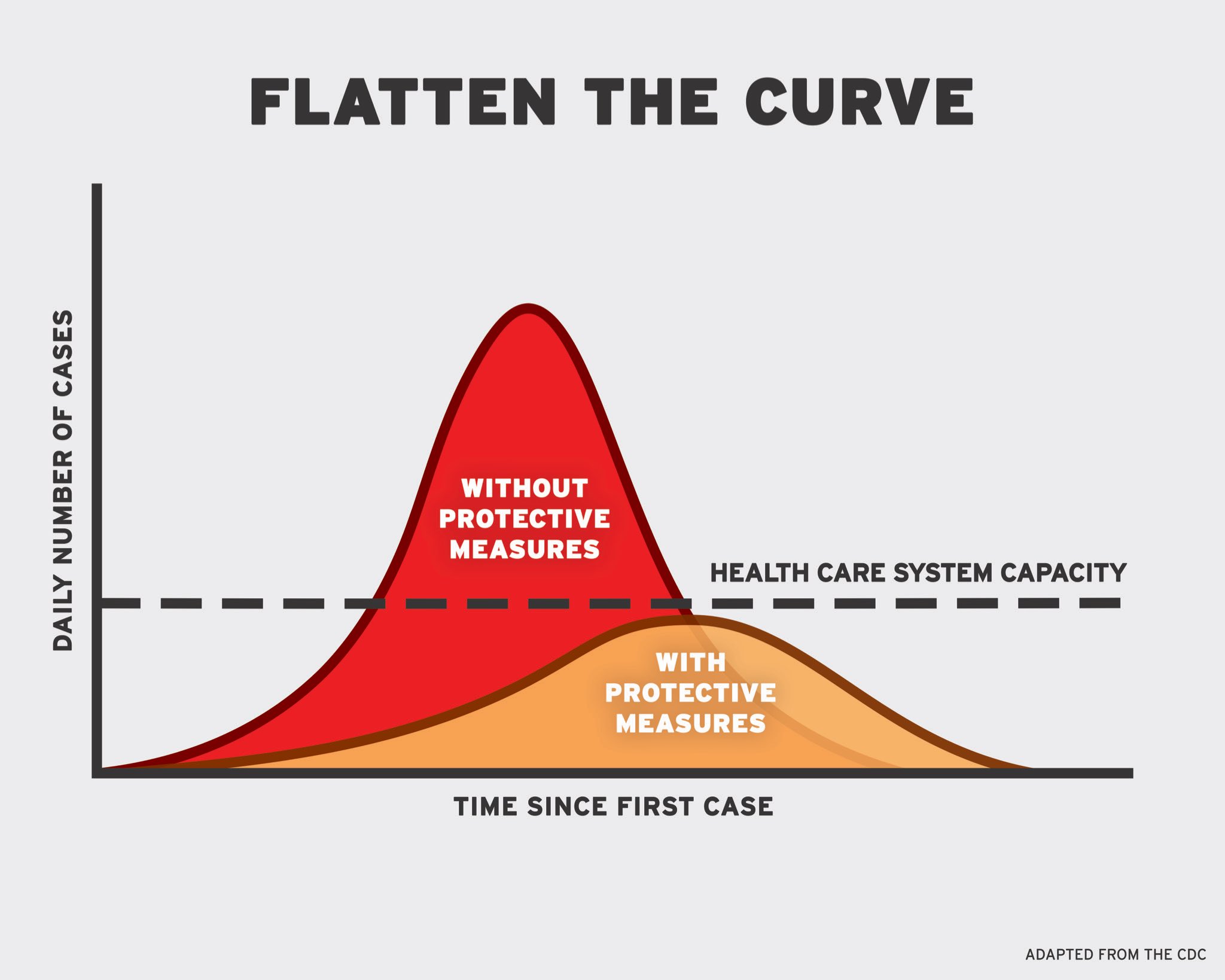}
        \caption{Governor of Illinois}
        \label{fig:illinois_ftc}
    \end{subfigure}
    
    \caption{The ``Flatten the Curve'' variants with more than 300 retweets during March 2020.}

\end{figure}

\section{Introduction}
Visualizations allow us to quickly convey a complex concept in a more succinct form.
% Understanding the spread of images has various uses. 
%Media may be interested in discovering copyright infringement, but they may also want to analyze the popularity and extension of their intellectual property. Additionally, d
Discovering the spread of an image online can be useful for understanding the reuse of scientific visualizations by the public. In this paper demonstrating preliminary results from our exploration of image spread, we use the ``Flatten the Curve'' (\textbf{FTC}) visualization as a case study for analyzing how one might discover the spread and reach of a particular image.

The \emph{New York Times} establishes a history for FTC and its association with COVID-19 \cite{NYT_Roberts_2020}. Per that history, \emph{The Economist} published in article in February 2020 \cite{Economist_2020} describing the COVID-19 pandemic and the steps countries were taking to combat the SARS-CoV-2 virus. This article contained a picture of two curves (Figure \ref{fig:economist_ftc}) contrasting the number of disease cases in a population depending on whether or not they adopted protective measures like wearing masks or social distancing. Health analyst Drew Harris \cite{Harris_Tweet_2020} read this article and was inspired to create Figure \ref{fig:harris_ftc}, adding a horizontal line demonstrating how the capacity of the health system is exceeded when the populace fails to adopt protective measures.

The phrase describing FTC was adopted as a Twitter hashtag \emph{\#FlattenTheCurve} by February 28, 2020. By conducting an advanced Twitter search of the \emph{top} (as opposed to \emph{latest}) \emph{\#FlattenTheCurve} tweets from March 2020 and manually identifying variants of this image with 300 or more retweets, we discovered three more variants, shown in Figures \ref{fig:lancet_ftc} - \ref{fig:illinois_ftc}. Thus, with our five variants of this image, we want to answer the following research questions:

\begin{itemize}
    \item \textbf{RQ1} What methods exists to help us explore the spread of these image variants and what are their limitations?
    \item \textbf{RQ2} Based on these methods, which FTC variant was reused the most and which had the most reach?
\end{itemize}

To understand the spread of these image variants, we leverage three different information channels: reverse image search engines, social media, and web archives. Through \textbf{reverse image search engines}, we submit each image variant as a \textbf{query image} and receive a set of corresponding \textbf{page URLs} for web documents that contain that image. This helps us understand the reuse of a given image. We then submit those page URLs to social media to understand when and how often they were shared online. Finally, we did the same for web archives to understand how often they were preserved, applying acts of preservation as a proxy for popularity. We find that the Harris variant (Figure \ref{fig:harris_ftc}) has the most reuse and preservation while the Lancet variant (Figure \ref{fig:lancet_ftc}) inspired a lot of social media chatter later in the pandemic. When building a digital library of web images, metadata about the use of those images may be needed. Our main contribution is demonstrating how the page URLs for documents containing images can serve as a simple and effective proxy for understanding image spread.

\section{Methodology}

We examined three different channels to quantify image spread: reverse image search engines, social media, and web archives. Because \emph{The Economist} article was published late in February 2022 and Harris created his variant on February 27, 2022, we manually conducted a Twitter Advanced Search for the top tweets that contained the hashtag \emph{\#FlattenTheCurve} posted between March 1 and 31 of 2020. We conducted this search in January 2023. We saved all images with 300 or more retweets because we are interested in the most popular variants of the FTC visualization. This gave us our base set of image variants to study.

% At the time, Twitter's search would only provide results for one day at a time. For example, to view tweets from March 1, 2020, we had to place bounds of March 1 and March 2 and then repeat for each of the 31 days in March. We assume we could not get all results in one search because Twitter has become unstable since much of the staff had been released \cite{cnbc_2023}. 

% With reverse image search, a query image, and not text, is the query and the search engine returns two types of results: images similar to the query image and pages containing the query image.
We then submitted each of these images to scrapers that we had developed for the reverse image search engines of Baidu, Bing, and Google.  Two types of results were available: those for pages containing the image, and those for images similar to our variant. We instructed our scrapers to save the results for pages containing the image. Our scrapers recorded the page URL of each page that contained the query image and saved the thumbnail image returned by the search engine for comparison. We manually reviewed all thumbnail images to ensure that they matched the corresponding query image, accepting images that were cropped or resized as the same image.

These results give us one perspective, that of search engines. We note that the reverse image search capability, while incredibly useful, does not give us a view over time, only the view of each search engine at the time of our query in January 2023. Social media can help us understand which FTC variant had the most reach. Unfortunately, social media platforms have yet to introduce reverse image search. Instead, we relied on the page URLs scraped from our search engine results as a proxy. To acquire social media results, we employed \emph{SNScrape} \cite{SNScrape}, a tool that can conduct text searches of Twitter and Reddit posts. We supplied each page URL as a text query and recorded the data about each post or comment that returned.

There are problems with this method. Sometimes an FTC variant was used in a web document for a limited period and then removed in favor of new content, related to \textbf{content drift} \cite{zittrain_albert_lessig_2014, klein_2014, jones_2016, zittrain_2021}. This often happens with news sites where a featured story contains the image on the front page of the site or a page related to a topic, which is then shared. Later a new featured story with a new image takes its place. To limit the effect of this, we restricted our results to only include social media posts after January 30, 2020; the date the the World Health Organization declared that COVID-19 was a world health emergency.

We address reuse with our search engine results. We address popularity through our social media results. Another proxy for popularity is measuring acts of preservation. Our final channel of image spread information is web archives. Searching one web archive would only give us a partial picture of the spread of an image. Memgator \cite{memgator_2016} leverages the Memento Protocol \cite{Jones2021} to help us search multiple web archives at once. Most web archives do not support reverse image search. Thus, again we relied on page URLs as a proxy for our images. To mitigate the content drift issues noted above, we provided the same date (January 30, 2020) as a lower bound for when each page was archived. In this paper the term \textbf{memento} refers to a single archived web page. Because the same page may be archived multiple times, a single page URL may have multiple mementos.

\begin{figure}
    \centering
    \includegraphics[width=0.45\textwidth]{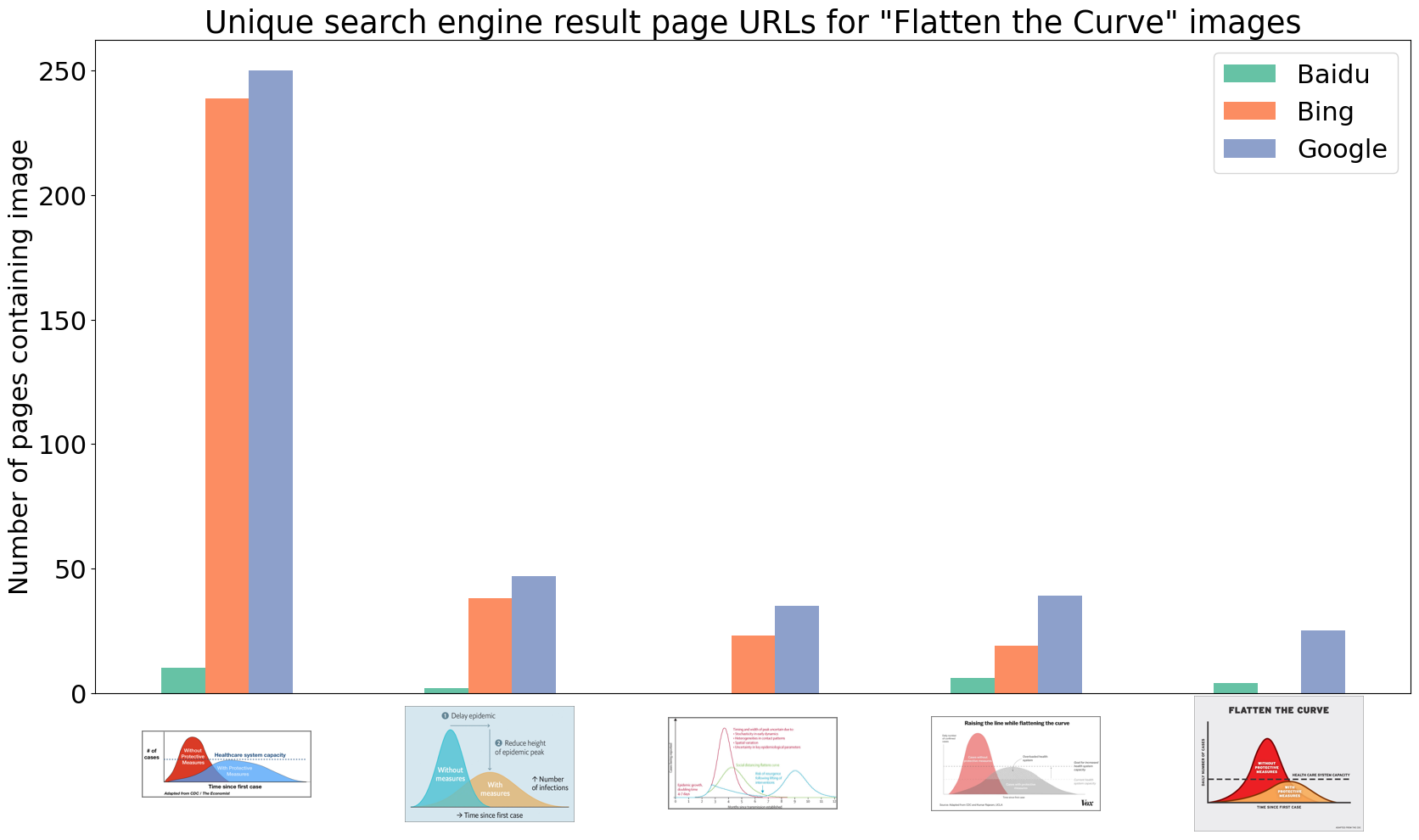}
    \caption{The number unique page URLs returned by each search engine for each query image, showing the Harris (Figure \ref{fig:harris_ftc}) variant as the most reused.}
    \label{fig:search_results}

    % \centering
    % \includegraphics[width=0.45\textwidth]{tld_spread.png}
    % \caption{The spread of domain suffixes among the page URLs for each image among search results.}
    % \label{fig:tld_spread}

\end{figure}

\begin{figure}

    \centering
    \includegraphics[width=0.5\textwidth]{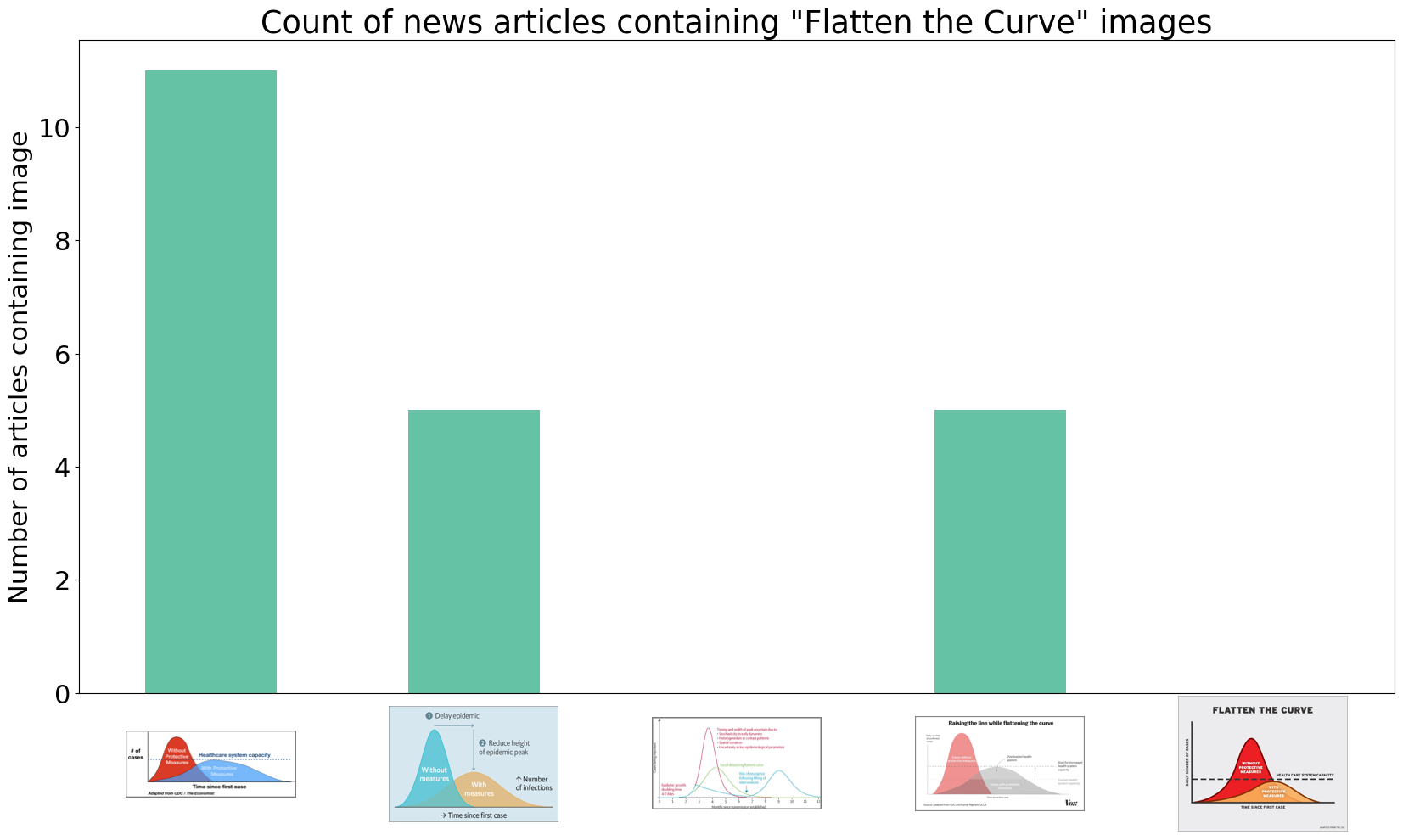}
    \caption{The number of news articles for each query image, as returned in the search engine results.}
    \label{fig:news_articles}
    
\end{figure}

\begin{figure}
    \centering
    \includegraphics[width=0.45\textwidth]{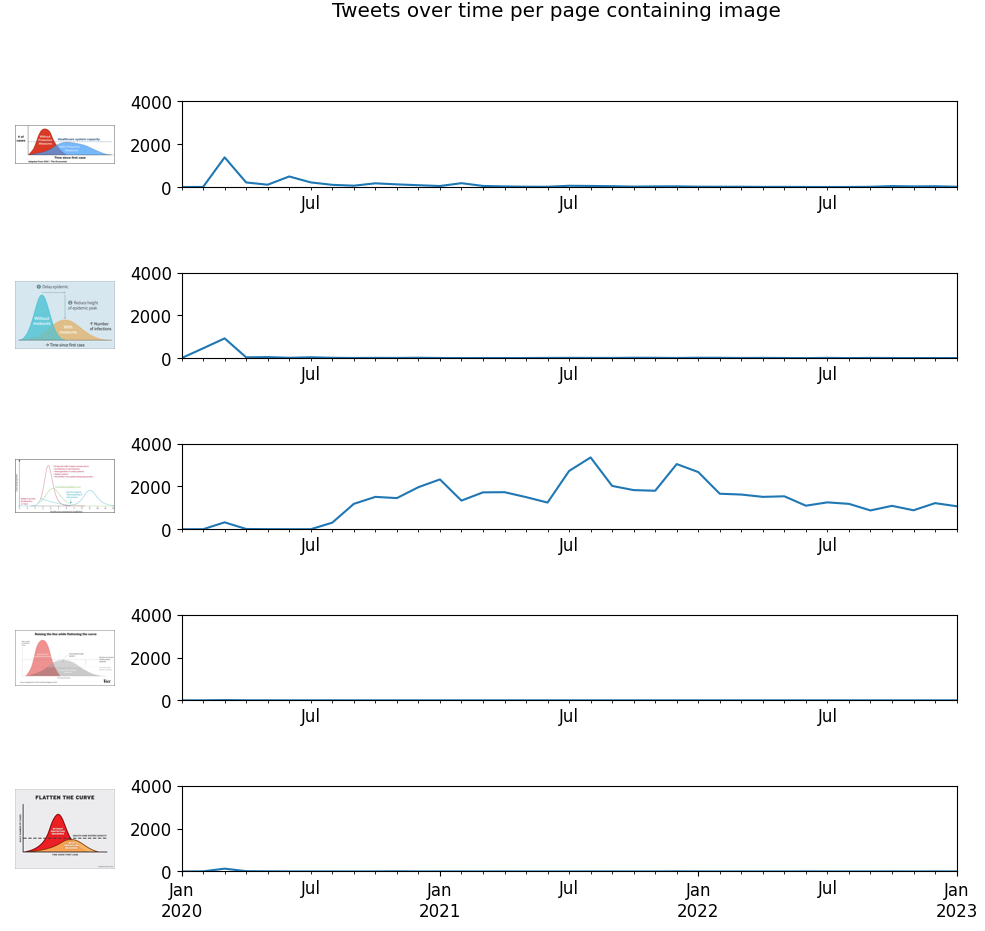}
    \caption{The history of Tweets sharing pages containing each image, showing the Lancet (Figure \ref{fig:lancet_ftc}) variant having the most activity.}
    \label{fig:tweets_over_time}

    \centering
    \includegraphics[width=0.45\textwidth]{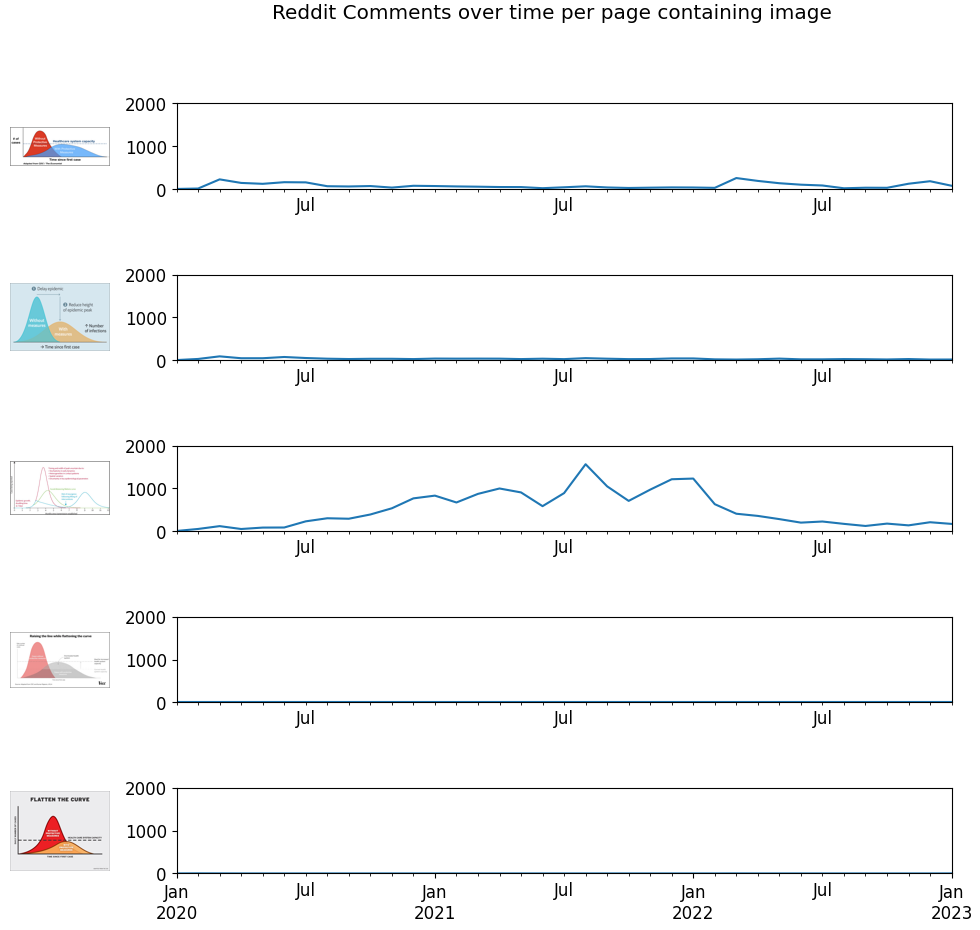}
    \caption{The history of Reddit comments sharing pages containing each image a comment on Reddit, showing the Lancet (Figure \ref{fig:lancet_ftc}) variant having the most activity.}
    \label{fig:reddit_over_time}

\end{figure}

\begin{figure}

    \centering
    \includegraphics[width=0.5\textwidth]{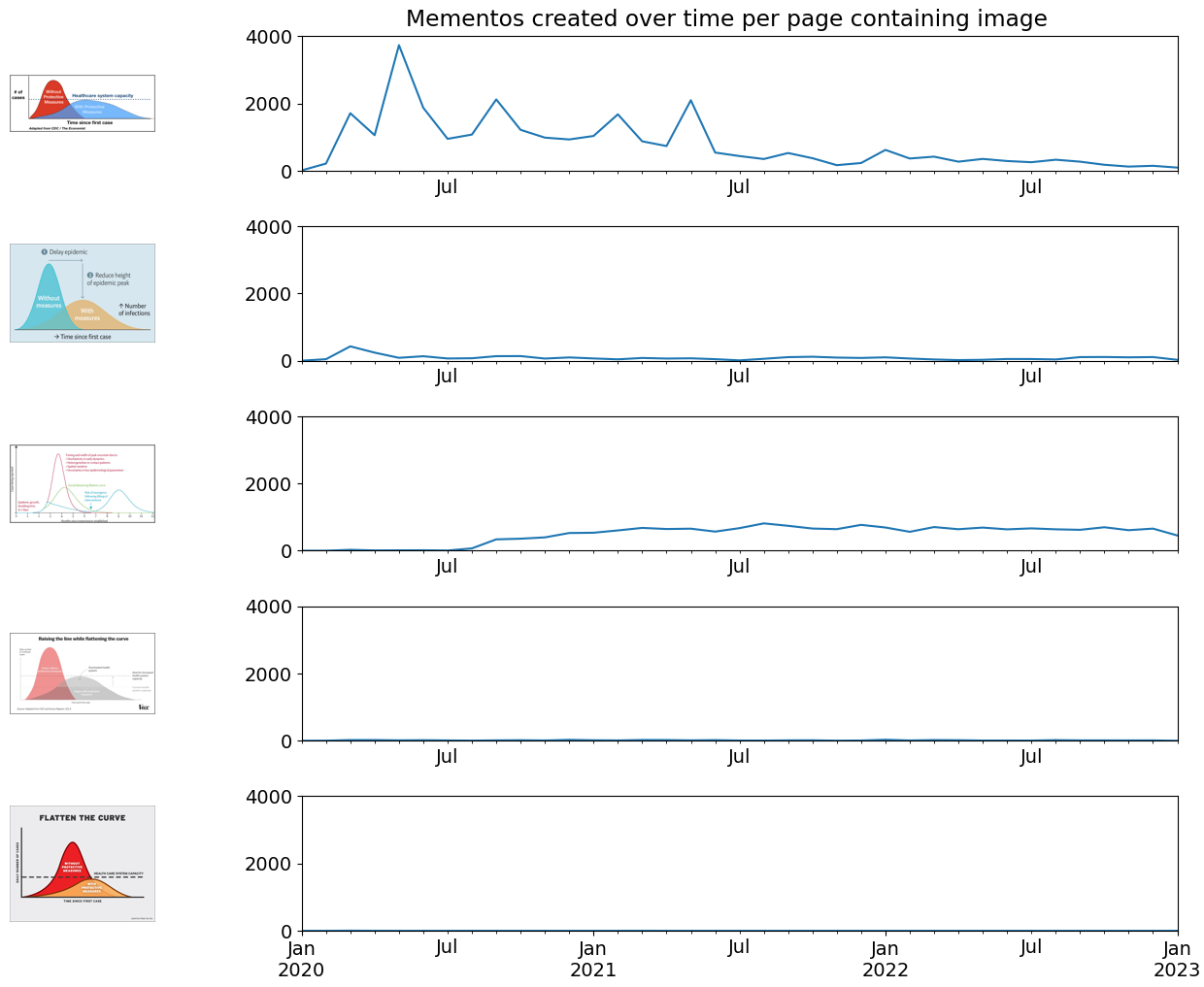}
    \caption{The history of memento creation for each page containing our Flatten the Curve images, showing the Harris (Figure \ref{fig:harris_ftc}) variant as the most preserved.}
    \label{fig:mementos_over_time}
    
\end{figure}

\section{Results and Discussion}

Figure \ref{fig:search_results} shows that the Harris variant (Figure \ref{fig:harris_ftc}) is the most widely discovered by the search engines. We see that the Economist image (Figure \ref{fig:economist_ftc}) while published earlier, is clearly less widely used in web documents as of January 2023. Other variants show far less use. 
% Figure \ref{fig:tld_spread} visualizes the domain suffixes of the pages containing each image. Here we clearly see that the Harris version is published in the largest diversity of pages, with the Economist image a distant second. 
Shown in Figure \ref{fig:news_articles}, we compared every page URL to the US News Domains list compiled by Robertson et al. \cite{news_domains_2018} and discovered that the Harris version is used in more than 10 current new articles.

%Examining social media provides a different picture. 
In Figure \ref{fig:tweets_over_time} we see each image variant along with how often an article containing them was shared on Twitter. Here we see that the Harris variant was shared early on in the pandemic while pages containing \emph{the Lancet} variant enjoyed a lot of activity from July 2020 onward. Figure \ref{fig:reddit_over_time} shows a similar pattern. We do not have the dates of retweets and SNScrape did not return Reddit posts prior to November 2022, so we are unable to provide  information about how well these images may have reached audiences through those methods. Web archives show the Harris variant again as the most popular. In Figure \ref{fig:mementos_over_time}, pages containing the Harris variant were heavily archived well into July of 2021.

Based on these results, the Harris variant was reused the most, but articles with \emph{the Lancet} variant engendered the most discussion and thus may have had more reach later in the pandemic. It is possible that both served different purposes. The Harris variant is very simple and was used to convince readers that the sacrifices they were making were worthwhile. Perhaps this is why it is more popular on social media toward the beginning of the pandemic. It is part concept visualization and part marketing. \emph{The Lancet} variant, however, is more complex and demonstrates additional concepts related to not only initially flattening the curve, but also illustrates the risks of resurgence should the populace stop engaging in protective measures. The resurgence peak in \emph{the Lancet} variant's curve occurs at nine months after the pandemic starts, which may correspond to the increase in sharing on both Reddit and Twitter as users are actively discussing the benefits and risks of relaxing protective measures.

\section{Related Work}

Choi \cite{choi_2010} examined how college students searched for images using text queries. Pu \cite{pu_2008} sought to understand the nature of failed image queries. Wang et al. \cite{wang_2007} explored better methods for displaying image search results. Several studies compare the effectiveness of different reverse image search platforms \cite{nieuwenhuysen2013search, kelly_2015, nieuwenhuysen2019finding, bitirim_2020, jones_2022}. Other studies seek to improve reverse image search \cite{gaillard2017large, gaillard2018, ARAUJO201835, 9339350, 9321037}. Zampoglou et al. \cite{zampoglou_2017} evaluated algorithms for identifying image forgeries on the web. We are not trying to analyze or improve image search as much as we are leveraging the tools that currently exist.

We are applying reverse image search as one of our information channels. It has been used in the past to study misinformation \cite{mci/Askinadze2017, aprin_2022, dhanvi_2022}, compare dermatology symptoms \cite{10.1001/jamadermatol.2016.2096, JIA20211415, SHARIFZADEH2021202}, understand political movements \cite{curran_2022, zahorova_2018}, create datasets \cite{10.1117/12.2228505}, create image captions \cite{10.1145/3173574.3174092}, and analyze biological protocols \cite{mamrosh_2015}. We differ by applying it using scrapers at scale to acquire thousands of page URLs which we then apply to the channels of social media and web archives. Our reverse image searches are similar to the work of Thompson \& Reilly \cite{thompson_2017} who suggest that reverse image search can help us understand image reuse within library collections.

%Jackson \cite{jackson_2012} applied web archives to study the changes of file formats on the web over time. 
Similar to Nwala et al. \cite{nwala_2021} we use acts of preservation as a proxy for document popularity. We are not, however evaluating the quality of URLs shared on social media. Many studies have evaluated images on social media \cite{wang_2015, kharroub_2016, Garimella_2016, neumayer_2018}, as well as built models of image features to predict popularity \cite{Niu_2012, Almgren_2016}. We are not analyzing the images as much as evaluating the URLs of documents that contain them. We are also trying to understand a viral image after the fact rather than trying to predict its virality.

\section{Future Work}

These are the results of a preliminary study into the spread of images online. For the specific search results that we analyzed, we were forced to use the page URLs as proxies because reverse image search engines do not return the image URLs in their results. Instead each search engine only provides image thumbnails. To acquire the image URLs, we would load the document behind each page URL and analyze every image using techniques such as VisHash \cite{oyen_2021}, pHash \cite{zauner_implementation_2010}, or SSIM \cite{zhou_2004} to find the image that corresponds to the query image. We could then supply these image URLs to web archives to see how often the images themselves were archived. We are developing tools to perform this action at scale.

Unfortunately, social media users do not often share image URLs, instead favoring page URLs, making this less useful for social media. We would acquire social media posts key to the time period and analyze their content. Using SNScrape, we could acquire all posts or comments related to a specific term or hashtag on either Twitter or Reddit. From there we could analyze the images associated with each post using the tools mentioned above. To begin, we would develop a list of search terms germane to our area of study.

As noted above with content drift, it is possible that some of these pages only contained an FTC variant for a short period of time. To mitigate this, inspired by document similarity work by SalahEldeen et al. \cite{salaheldeen_2015} and Jones et al. \cite{jones_2016}, we attempted to leverage mementos of these pages to verify that they did contain FTC variants during the time period that they were shared on social media. We discovered additional challenges to applying this method with images. Many pages were not preserved close enough to the date they were shared on social media, or they suffered from memento damage \cite{brunelle_2015}, a phenomena where web documents are preserved without some of their images or the code necessary to load them. We are exploring ways to work around this damage by analyzing the HTML for specific image URLs.

\section{Conclusion}

Visualizations can often convey complex concepts more briefly than text explanations. Understanding the spread of visualizations can be beneficial to understanding how the public consumes science. The ``Flatten the Curve'' (FTC) visualization was key to understanding the benefit of protective measures during the COVID-19 pandemic. It spawned several popular variants. In this paper, we use FTC as a case study to explore how one might discover the spread and reach of a particular image.

We addressed RQ1 by exploring how one might apply reverse image search to first find the web documents that currently contain an image. We found that the Harris variant was reused in the most pages based on results from Baidu, Bing, and Google. We then submitted the URLs of these pages to social media to see how often and when they were shared. We discovered that \emph{the Lancet} variant was shared heavily on social media, potentially during a time when protective measures were waning. Through acts of preservation, we found the Harris variant to be more popular during the beginning of the pandemic.

We encountered limitations with respect to time. Using reverse image search provides us only with a current view of which pages contain a given image. Social media platforms and web archives do not typically have reverse image search capability. This leads us to use a proxy to estimate the reach of each image. Because we could search Twitter and Reddit for URLs, we used the page URL for the document containing the image as this proxy. This is potentially problematic because a web document may contain the image today, but may not have in the past, even though the document maintained the same URL. We mitigated this content drift by bounding the time period for posts and web archives in our dataset, but we are researching better solutions.

Per RQ2, our case study demonstrated that the Harris variant continues to be the most popular in terms of published work, but the Lancet variant created a lot of conversation over time. Iconic images such as Flatten the Curve have a daily impact on our lives, shaping policy and public action. While building a digital library of images, capturing metrics to understand the spread of such images may be as important as understanding the spread SARS-CoV-2 itself.

\bibliographystyle{ACM-Reference-Format}
\bibliography{refs}

\end{document}